\documentclass[]{spie}  

\usepackage{amsmath,amsfonts,amssymb}
\usepackage{booktabs}
\usepackage{siunitx}
\DeclareMathOperator*{\argmin}{arg\,min}
\usepackage{graphicx}
\usepackage[colorlinks=true, allcolors=blue]{hyperref}

\title{ESCAPE: Efficient Synthesis of Calibrations for Adaptive optics through Pseudo-synthetic and Empirical methods}

\author[a,b]{Jacob Taylor}
\author[a,c]{Robin Swanson}
\author[g,h]{Parker Levesque}
\author[e]{Masen Lamb}
\author[f]{Amali Vaz}
\author[f]{Manny Montoya}
\author[f]{Andrew Gardner}
\author[f]{Katie M. Morzinski}
\author[a,b]{Suresh Sivanandam}

\affil[a]{Dunlap Institute for Astronomy and Astrophysics, University of Toronto, 50 St George St, Toronto, ON M5S 3H4, Canada}
\affil[b]{David A. Dunlap Department of Astronomy and Astrophysics, University of Toronto, 50 St George St, Toronto, ON M5S 3H4, Canada}
\affil[c]{Department of Computer Science, University of Toronto, 40 St George St, Toronto, ON M5S 2E4, Canada}
\affil[e]{Gemini Observatory (United States), 670 N. A’ohoku Place, Hilo, United States}
\affil[f]{Steward Observatory, University of Arizona, 933 N Cherry Ave, Tucson, AZ 85719, United States}
\affil[g]{Department of Physics, Université de Montréal, 1375 Avenue Thérèse-Lavoie-Roux, Montréal, QC, H2V 0B3}
\affil[h]{Mila - Quebec Artificial Intelligence Institute, Université de Montréal, 6666 St-Urbain St, Montreal, QC H2S 3H1, Canada}

\authorinfo{Further author information: (Send correspondence to J.T.)\\J.T.: E-mail: jacob.taylor@mail.utoronto.ca}

\begin{document} 
\newcommand{\defIM}{\mathbf{D}}

\newcommand{\defRM}{\mathbf{R}}

\newcommand{\defSlopes}{\vec{\mathbf{s}}}

\newcommand{\defCommands}{\vec{\mathbf{c}}}

\newcommand{\defWave}{\mathbf{W}}

\newcommand{\defMisreg}{\vec{\alpha}}
\maketitle

\begin{abstract}
With the commissioning of the refurbished adaptive secondary mirror (ASM) for the 6.5-meter MMT Observatory under way, special consideration had to be made to properly calibrate the mirror response functions to generate an interaction matrix (IM).
The commissioning of the ASM is part of the MMT Adaptive optics exoPlanet characterization System (MAPS) upgrade the observatory's legacy adaptive optics (AO) system.
Unlike most AO systems, MAPS employs a convex ASM which prevents the introduction of a calibration source capable of simultaneously illuminating its ASM and wavefront sensor (WFS). 
This makes calibration of the AO system a significant hurdle in commissioning. 
To address this, we have employed a hybrid calibration strategy we call the Efficient Synthesis of Calibrations for Adaptive Optics through Pseudo-synthetic and Empirical methods (ESCAPE). 
ESCAPE combines the DO-CRIME on-sky calibration method with the SPRINT method for computing pseudo-synthetic calibration matrices. 
To monitor quasi-static system change, the ESCAPE methodology rapidly and continuously generates pseudo-synthetic calibration matrices using continual empirical feedback in either open or closed-loop. 
In addition, by measuring the current IM in the background while in close-loop, we are also able to measure the optical gains for pyramid wavefront sensor (PyWFS) systems. 
In this paper, we will provide the mathematical foundation of the ESCAPE calibration strategy and on-sky results from its application in calibrating the MMT Observatory's ASM. 
Additionally, we will showcase the validation of our approach from our AO testbed and share preliminary on-sky results from MMT. 
\end{abstract}


\section{MOTIVATION}
\label{sec:motivation}  

In adaptive optics (AO), the interaction matrix (IM) is a critical component that defines the relationship between the system's wavefront sensor (WFS) and its wavefront corrector (WFC), nominally a deformable mirror (DM).
This matrix is determined during system calibration by applying a set of known orthonormal shapes to the corrector and recording the sensor's corresponding responses.
Under the assumption that the mapping between the injected WFC disturbance and corresponding WFS response is linear, the resulting WFS responses define the linear transformation between the WFS and WFC, or equivalently, they define the IM. 
An accurate IM is essential for effective wavefront correction and greatly impacts system performance.
While this calibration scheme is applied almost ubiquitously across AO systems, it has one fundamental assumption; the WFS and WFC must be simultaneously illuminated by an aberration-free source.
Conventional AO systems include an internal calibration source which is co-aligned with the telescope beam.
This source is used to perform several alignment and calibration tasks, including the acquisition of the IM.
These tasks are typically done during the daytime when the system is otherwise unused, thereby not impacting the system's overall efficiency.
This paper is concerned with how the nominal calibration methodology breaks down in the context of telescope designs where the inclusion of a  calibration source is impractical.

\begin{figure} [ht]
\begin{center}
\begin{tabular}{c}
\includegraphics[trim={0cm 0cm 3cm 0cm},clip, width=0.95\textwidth]{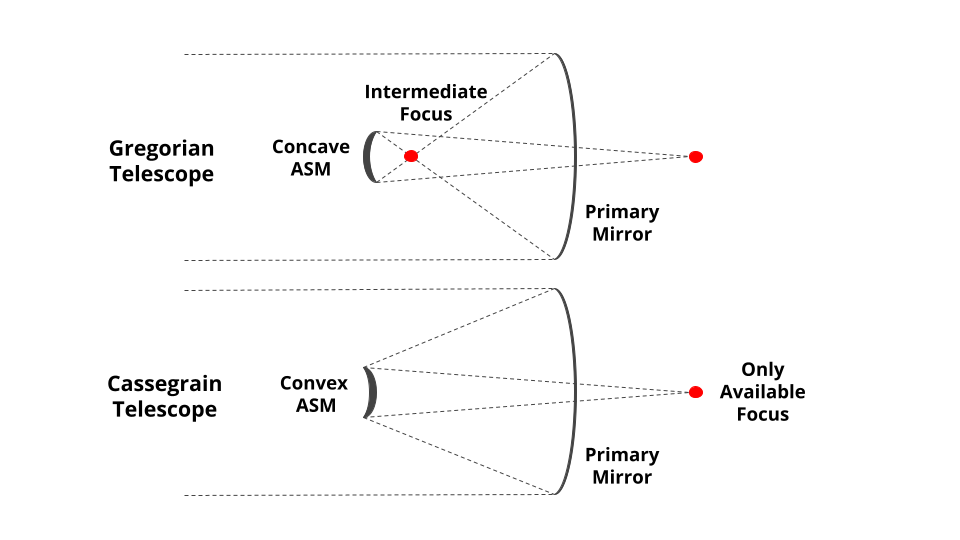}
\end{tabular}
\end{center}
\caption[Concave vs Convex ASMs] 
{A schematic illustrating the difference in available focal planes between a Gregorian telescope (concave secondary) and a Cassegrain telescope (convex secondary). In the case of the Gregorian telescope, light comes to a focus before reflecting off of the secondary mirror. This intermediate focal plane allows for the introduction of a calibration source which can illuminate the secondary mirror. This is in contrast to the Cassegrain design which employs a convex secondary mirror. In this case, there is no focal plane before the secondary mirror in the optical chain, and therefore, the introduction of a calibration source which illuminates the secondary is effectively impossible.}
\label{fig:concaveVconvex}
\end{figure} 

Several modern observatories --- the MMT, the Large Binocular Telescope (LBT), the Very Large Telescope (VLT), among others --- have opted to have their secondary mirror act as their AO system's WFC, a so-called Adaptive Secondary Mirror (ASM).
This class of WFC can be more generally described as `pre-focal' DMs, indicating that the DM is located prior to the telescope's main focal plane.
Most notably, the designs of both the Extremely Large Telescope (ELT) and the Giant Magellan Telescope (GMT) include large, segmented pre-focal DMs.
For this paper, we will use the term ASM to refer to all pre-focal DM configurations.
The ASM approach has several benefits, namely, ASM designs reduce the total number of optical elements in the optical chain before the science instrument, thereby increasing throughput and lowering the observed infrared background.
ASMs also allow for correction over a much wider field than their post-focal counterparts, making them particularly well-suited for techniques like ground-layer adaptive optics (GLAO). 
While employing an ASM can have substantial scientific improvements, it can also increase the cost and complexity of the AO system itself.
While the optical designs of ASM-equipped AO systems will requires less total components, in practice, the added complexity of an ASM can be a significant hurdle.
ASMs are necessarily much larger than their post-focal plane counterparts, often reaching sizes on the order of $\approx 1$ meter or more.
This has large implications for ASMs actuator design, power \& thermal requirements, generally resulting in a much more expensive and complex AO system.
Most important for this work, moving the WFC further up the optical chain has a critical impact on the placement of the system's calibration source.
For designs which employ a concave ASM such as the Large Binocular Telescope (LBT), a calibration can be placed at the telescope's intermediate focus between its primary and secondary mirrors and brought into and out of the beam with the use of a robotic arm.
While this adds an additional moving component to the system design, it still remains practical to introduce a calibration source into the optical chain.
However, for telescopes like the MMT which employ a convex ASM, there is no available focal plane before the ASM where a calibration source could be injected into the system.
This makes jointly illuminating the WFC and WFS systems simultaneously with an aberration-free source effectively impossible.
Figure~\ref{fig:concaveVconvex} illustrates the difference between employing a concave ASM compared to a convex ASM.

The lack of calibration source for AO systems hosting pre-focal, convex ASMs has several compounding consequences for operations and AO control.
As discussed below, several of these complications demand a disproportional amount of an observatory's most precious resource --- sky-time.
First, all alignment tasks between the ASM and the WFS must be done using turbulent starlight.
From an operations perspective, this can have serious implications since some fraction of sky-time must be allocated for alignment tasks, which could otherwise be done during the day or in a lab.
Furthermore, starlight is faint compared to a calibration source, which makes any manual alignment tasks more difficult.
Overall, systems with convex ASMs will require well devised alignment procedures in order to minimize wasted sky-time.
Second, several of the software debugging tasks which would normally be performed in a lab environment under idealized conditions must be done using starlight.
For example, tip-tilt offloading from the AO system to the telescope will require some calibration procedures and software.
Under normal circumstances, this task could be worked on during daytime calibrations, or in the lab; however, since this task requires a closed-loop AO system in order to undergo verification testing (which requires joint illumination), it relies on sky-time for debugging, as well as the existence of a valid IM.
This effect compounds across the large set of software tasks that comprise facility AO operations.
Lastly, and most important to this work, without access to a calibration source, traditional methods for computing the system's IM become impossible.

This calibration problem has been well known in the AO community since at least 2004\cite{MMT_ONSKY_IM_2004}, where MMT AO scientists first proposed methods for trying to measure an IM using starlight.
These methods were further refined with the introduction of the sinusoidal modulation technique by Esposito et al. (2006) \cite{MODULATION_TECHNIQUE}, which proposed modulating the amplitude of modes applied on the DM at a known frequency and then recovering the WFS response by demodulating the slope signal.
That same year, Oberti et al. \cite{pseudo-synthetic_2006} proposed the first pseudo-synthetic approach for improving simulated IMs using empirical information about the system, a methodology that would later become incredibly influential.
These formative ideas served as a framework for testing on-sky IM methodologies on the then newly commissioned First Light Adaptive Optics system (FLAO) at the Large Binocular Telescope (LBT)\cite{LBT_ON_SKY}.
Since the LBT employs a concave ASM, scientists were able to compare their on-sky IMs to a nominal calibration source IM as well as an accurate synthetic model based on high quality empirical data.
With that said, Pinna et al. (2012)\cite{LBT_ON_SKY} presented the first high order on-sky interaction matrices taken using LBT's FLAO system utilising the sinusoidal modulation technique.
The authors were able to demonstrate a robust methodology for generating empirical on-sky interaction matrices, but conclude that a hybrid approach that combines their empirical methods with a pseudo-synthetic calibration is preferred.
The paper concludes, ``the two techniques can be combined in an efficient calibration strategy for high order systems, particularly those provided of convex adaptive secondary mirrors."
This testing on the LBT was conducted in the wake of the VLT adaptive optics facility (AOF) coming online which did not have access to a calibration source.
Kolb et al. (2012)\cite{VLT_CALIBRATION_STRATEGY} details the AOF's baseline for using a pseudo-synthetic calibration strategy to get an initial closed-loop.
While this pseudo-synthetic calibration methodology proved highly effective for closing the loop on relatively low order modes, for controlling higher order modes the AOF team implemented the sinusoidal modulation technique \cite{ERIS_CAL_BOOTSTRAP}, thereby realizing the hybrid ideal first proposed by Pinna et al.\cite{LBT_ON_SKY}

While the modulation-demodulation methodology (among others) has historically performed well, its main limitation is complexity; in order to reduce the amount of sky-time required to acquire the IM, several modes must be multiplexed at different frequencies which can be complex to implement, particularly as the number of modes grows.
Recently, Lai et al. (2020)\cite{DOCRIME_PAPER} have proposed the DO-CRIME method for rapid empirical on-sky interaction matrix evaluation by applying a series of randomized mirror commands and cross-correlating those commands with the observed slope signal.
Their methodology has a number of appealing features: the IM can be determined very quickly in both open and closed loop, the complexity of software implementation is low, and all modes can be measured simultaneously allowing for seamless application between system architectures.
These practicalities make the algorithm particularly appealing for facility-class systems where sky-time is often limited and costly. 
Along with advancements in empirical methods, new tools for generating pseudo-synthetic interaction matrices have also become available in recent years; namely, Heritier et al. (2021)\cite{SPRINT_PAPER} present a comprehensive methodology for compensating for \textit{mis-registrations} between simulated and empirical systems. 
The methodology, called SPRINT, assumes that a system's mis-registrations can be characterized by an affine transformation between the DM and the WFS.
This assumption reduces the complexity of the problem to only a handful of physically motivated parameters. 
The partial derivation of the IM with respect to each parameter is calculated and compiled into the \textit{meta sensitivity matrix}, which is then used along with an empirically determined IM to compute the optimal parameters of the affine transformation.

In this work, we will introduce the ESCAPE calibration technique which combines the DO-CRIME and SPRINT techniques into a powerful method for calibrating AO systems without access to a calibration source.
In addition, we will show how the DO-CRIME calibration technique is well-suited to systems utilizing a pyramid wavefront sensor (PyWFS) since its passive IM calculation can be used to extract optical gains during closed-loop control.
Furthermore, we will present lab results where we have tested the ESCAPE methodology on our optical bench.
Lastly, we will give an update on our testing of the ESCAPE methodology on-sky at the MMT. 

\section{Methodology}
\label{sec:methodology}

In their linear approximation, adaptive optics systems are repeatedly converting WFS response signals into a corresponding set of WFC commands, summarized by the equation:
\begin{equation}
    \defCommands = \defRM \cdot \defSlopes
\end{equation}
where $\defSlopes$ is the \textit{slope vector} measured by the wavefront sensor representing the current atmospheric aberrations, $\defCommands$ is the \textit{command vector} which contains the DM command that would correct for the observed atmospheric aberrations, and $\defRM$ is a linear transformation called the \textit{reconstruction matrix} which maps slope vectors to their corresponding command vector.
In practice, the matrix $\defRM$ is difficult to empirically construct and is therefore approximated as the pseudo inverse of the \textit{interaction matrix}, $\defIM$.
The interaction matrix, or IM, is defined by the linear transformation that solves the inverse problem to the reconstruction matrix, namely:
\begin{equation} \label{ref:s_IM_dot_c}
    \defSlopes = \defIM \cdot \defCommands
\end{equation}
The IM can be empirically constructed by applying a series of orthogonal basis command vectors to the DM while illuminating with an unaberrated point-source and recording the WFS slopes.
Mathematically, to produce an IM with $N$ control modes: 
\begin{equation}
    \defIM = \begin{bmatrix}\frac{\defSlopes(\alpha\epsilon_1) - \defSlopes(-\alpha\epsilon_1)}{2\alpha},& \frac{\defSlopes(\alpha\epsilon_2) - \defSlopes(-\alpha\epsilon_2)}{2\alpha}, & \dots, & \frac{\defSlopes(\alpha\epsilon_N) - \defSlopes(-\alpha\epsilon_N)}{2\alpha}\end{bmatrix}
\end{equation}
where each $\defSlopes(\alpha\epsilon_i)$ is a column of the interaction matrix and represents the slope signal observed by the WFS for a given basis command $\epsilon_i$ scaled by $\alpha$.
In practice, $\alpha$ is chosen to be small enough to maintain the linearity of the WFS, and each WFS response is averaged over many frames to improve the signal-to-noise of the measurement.
Furthermore, the average between a positive and negative basis perturbation are taken to improve robustness and remove any sensor offset; because of this, the method is often called the `push-pull' method.

The chosen control basis is either \textit{zonal}, where each actuator on the DM is treated as a basis vector, or more often \textit{modal}, where the orthogonal basis vectors are continuous shapes that span the mirror.
Once the IM has been constructed, the reconstruction matrix is then typically computed using a singular value decomposition (SVD) so that modes which are poorly sensed by the wavefront sensor can be filtered out. 
Mathematically, the IM is decomposed as the product of orthonormal matrices, $\mathbf{U}$ and $\mathbf{V}$, and a diagonal matrix, $\mathbf{\Sigma}$, such that the diagonal entries of $\mathbf{\Sigma}$ are the eigenvalues of $\defIM$:
\begin{equation}
    \defIM = \mathbf{U} \cdot \mathbf{\Sigma} \cdot \mathbf{V}^*
\end{equation}
under this construction, the reconstruction matrix can be computed as:
\begin{equation}
    \defRM = \defIM^+ = \mathbf{V} \cdot \mathbf{\Sigma^+(\mathbf{\phi})} \cdot \mathbf{U}^*
\end{equation}
where $\mathbf{\phi}$ is an integer scalar controlling how many eigenmodes are included in the inversion and $\defIM^+$ is the Moore-Penrose inverse of $\defIM$.

The methodology described above is the dominant calibration scheme used for AO systems, however, a more nuanced explanation can be found in Meimon et al. (2015)\cite{AO_Calibration_Indepth}. 
Moving forward, We will refer to IM constructed this way either as \textit{standard interaction matrices} or as the baseline IM.
however this methodology relied entirely on illumination from an unaberrated point-source.
For AO system without access to such a calibration source, they must resort to using astronomical sources for their calibration.
This is problematic for the standard IM acquisition methodology since the induced WFS response signal intorduced by our perturbations will be much smaller than the induced signal from the atmospheric turbulence.
As described in Heritier et al. (2018)\cite{heritier_2018}, the two methods to reduce the noise are to either integrate long enough that the atmospheric signal averages out, or take fast push-pull measurements in order to effectively freeze the atmosphere between measurements. 

The following subsections explore alternative methods for computing the IM of an AO system, particularly in scenarios where the standard methodology fails. 


\subsection{DO-CRIME: Dynamic On-sky Covariance Random Interaction Matrix Evaluation}
\label{sec:DO-CRIME}

Following Lai et al. (2020)\cite{DOCRIME_PAPER}, we describe the how the IM is determined using the DO-CRIME method (hereafter called the \textit{DO-CRIME matrix}).
The method builds upon existing methods for generating on-sky IMs and is particularly effective due to its speed and relative simplicity.
We encourage the reader to reference the original paper for a more detailed description of the method.

Starting with Equation~\ref{ref:s_IM_dot_c}, we can add the time dependence of the system as the atmospheric conditions evolve:
\begin{equation}
    \defSlopes(t) = \defIM \cdot \defCommands(t)
\end{equation}
To construct a DO-CRIME matrix, a set of uniformly distributed randomized command vectors, $ \defSlopes_{\zeta}(t)$, are applied to the deformable mirror.

In open-loop, this process is summarized by the equality:
\begin{equation} 
    \defSlopes_{\zeta}(t) + \defSlopes_{a}(t) = \defIM \cdot \defCommands_{\zeta}(t) +  \defIM \cdot \defCommands_{a}(t)
\end{equation}

where $\defSlopes_{a}(t)$ and $\defCommands_{a}(t)$ signify the atmospheric aberration contribution. 
To generate an IM, we consider the time average of the above equation when multiplied by $\defCommands_{\zeta}^T(t)$:
\begin{equation} \label{eq:time_avg}
    \langle (\defSlopes_{\zeta} + \defSlopes_{a}) \cdot \defCommands_{\zeta}^T \rangle = \defIM \cdot  \langle (\defCommands_{\zeta} + \defCommands_{a} ) \cdot \defCommands_{\zeta}^T \rangle 
\end{equation}
Since the commands are randomized, we should expect no correlation between the randomized components and the atmospheric components. 
That is to say, given enough iterations we have: 
\begin{equation}
    \langle \defSlopes_{a} \cdot \defCommands_{\zeta}^T \rangle  = \langle \defCommands_{a} \cdot \defCommands_{\zeta}^T \rangle = 0 .
\end{equation}

We can then rearrange Equation~\ref{eq:time_avg} to derive an equation for $\defIM$:
\begin{equation}\label{eq:DO-CRIME_IM}
    \defIM =  \langle \defSlopes \cdot \defCommands_{\zeta}^T \rangle \langle \defCommands_{\zeta} \cdot \defCommands_{\zeta}^T \rangle ^ {-1}
\end{equation}

Equation~\ref{eq:DO-CRIME_IM} provides a practical and fast way to compute a system's IM with few assumptions; a DO-CRIME matrix is computed via a simple cross correlation between a history of randomized commands and their associated slopes.
Moreover, a DO-CRIME matrix can be computed in either open-loop, closed-loop, with a calibration source, or on-sky, making it an incredibly versatile calibration method.

\subsection{SPRINT: System Parameters Recurrent INvasive Tracking}
\label{sec:SPRINT}
In this section, we provide a summary of the SPRINT method following  Heritier et al. (2021) \cite{SPRINT_PAPER} and how our implementation differs slightly from the original.
We encourage the reader to reference the original for further details.
Following  Heritier et al. (2021)\cite{SPRINT_PAPER}, we introduce the idea of the \textit{registration} of an adaptive optics simulation. 
Since an AO system is comprised of physical components that are manually aligned, there exists a non-trivial abstraction of the physical state describing the positions and orientation of each component of the system with respect to one another. 
For well-aligned systems, this arbitrarily complex description of the system can be approximated as an affine transformation describing the coupled geometry of the DM and WFS, while assuming all other optics are ideal. 
The set of parameters describing this transformation, individually called \textit{misregistrations}, are collectively known as the registration of the system.
Furthermore, a simulated system is initially said to be \textit{unregistered} with respect to the physical system it is simulating.
A system's calibration matrices are highly sensitive to each of the registration parameters which typically prevents simulated calibrations of unregistered simulations from performing as well as their empirical counterparts when applied to their physical system. 


The registration of a system, denoted by a vector $\defMisreg$, is comprised of scalar quantities indicating the magnitude of each respective misregistration; in the SPRINT implementation the possible misregistration parameters are rotation ($\phi$), translation ($X$ and $Y$), and scaling (radial, $\rho$, and tangential, $\theta$):
\begin{equation}
    \defMisreg = \begin{bmatrix} \mathbf{\alpha_X}, & \mathbf{\alpha_Y}, &  \mathbf{\alpha_\phi}, & \mathbf{\alpha_\rho},  & \mathbf{\alpha_\theta}
    \end{bmatrix}
\end{equation}
Under this construction, we consider the first order expansion of the IM around a given system registration, $\defIM_{\alpha}$, which can be written as follows: 
\begin{equation}
    \mathbf{D_\alpha} = \mathbf{G}\left(\defIM_{\alpha_0}  + \sum_i \alpha_i \delta \mathbf{ D}_{\alpha_0} (\epsilon_i)\right)
\end{equation}
where $\defIM_{\alpha_0}$ is an unregistered IM, $\mathbf{G}$ is a diagonal matrix of scalar gains to account for the differential gain between the unregistered and registered systems, $\defMisreg$ is the misregistration vector, and $\mathbf{\delta D}_{\alpha_0} (\epsilon_i)$ is the sensitivity matrix corresponding to the $i^{th}$ misregistration. 
The sensitivity matrices $\delta \defIM_{\alpha_0}(\epsilon_i)$ are defined as the partial derivatives of the unregistered IM, $\defIM_{\alpha_0}$, with respect to the $i^{th}$ misregistration:
\begin{equation}
    \delta \defIM_{\alpha_0}(\epsilon_i) = \left( \dfrac{\defIM_{\alpha_0 + \epsilon_i} - \defIM_{\alpha_0 - \epsilon_i}}{2\epsilon_i} \right)
\end{equation}
and subsequently, the \textit{meta sensitivity matrix} is defined as the vector of all sensitivity matrices:
\begin{equation}
    \mathbf{\Lambda_{\alpha}} = \begin{bmatrix}
        \delta \defIM_{\alpha}(\epsilon_{X}), & \delta \defIM_{\alpha}(\epsilon_Y), & \delta \defIM_{\alpha}(\epsilon_\phi), & \dots
    \end{bmatrix}
\end{equation}

\begin{figure} [ht]
\begin{center}
\begin{tabular}{c}
\includegraphics[width=0.9\textwidth]{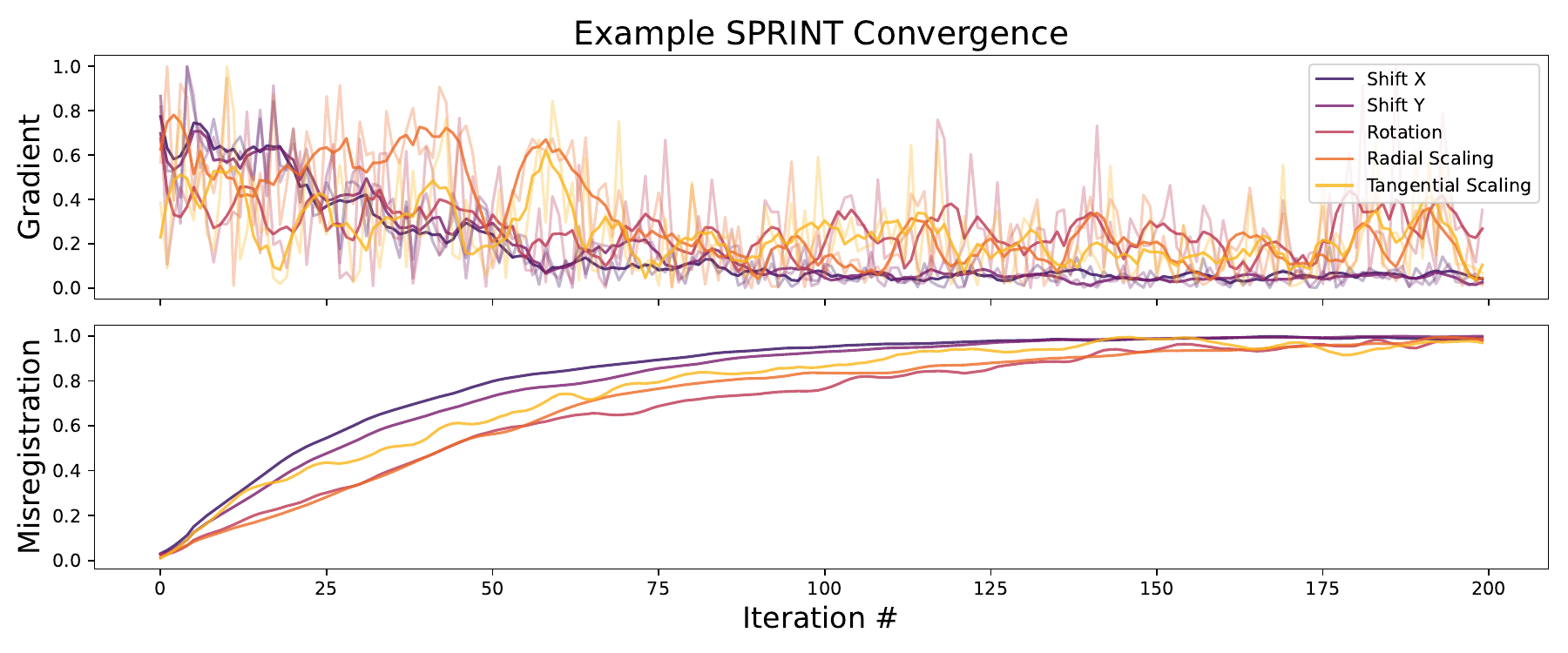}
\end{tabular}
\end{center}
\caption[SPRINT Convergence] 
{ \label{fig:convergence} An example of a typical misregistration convergence graph from our SPRINT implementation. (Top) The normalized magnitude of the misregistration gradient found for the $i^{th}$ iteration of the descent. (Bottom) The corresponding normalized misregistration vector to the gradients above. The plots have been normalized for easier comparison across parameters of different magnitudes and units. We see that the parameters slowly converge to a final misregistration vector, which we take to be our solution. Note that the noisier gradients are caused by there being little of that misregistration in our system, and therefore, appear relatively noisier once normalized.}
\end{figure}

The goal of the SPRINT procedure is to solve for the optimal misregistration vector, $\defMisreg^{\,*}$, and gain matrix, $\mathbf{G}$, in order to recover the registered IM for the system. 
If we are given an unbiased estimate of the true IM, $\hat{\mathbf{D_\alpha}}$ (e.g. an empirical on-sky IM), then the solution to this problem can be found by optimizing:
\begin{equation}\label{eq:argmin_full}
    \defMisreg^{\,*}= \argmin_{\mathbf{G},\defMisreg^{\,}} \| \hat{\mathbf{D}_\alpha} -\mathbf{G}\left(\mathbf{D_{\alpha_0}}   + \defMisreg^{\,}\mathbf{\Lambda_{\alpha_0}}\right)\|
\end{equation}
The SPRINT method implements an iterative solving paradigm which alternates between solving for the optical gain matrix, $\mathbf{G}$, and the misregistration vector, $\defMisreg$.
Alternatively, we present a simplified least squares solution by assuming that the optimal geometric transformation will be the same without accounting for mode dependent gain variations.
That is to say, we take the matrix, $\mathbf{G}$, to be adequately approximated by a single scalar gain, $g$.
For our envisioned use case, namely when the empirical IM fed into the misregistration solver is an on-sky IM taken in open-loop, there will be significant modally dependent optical gain differences between the synthetic and empirical matrices.
Under this condition, fitting for the modally dependent gain factors is not desired since the empirical matrix will not have the correct gains for closed-loop operations.
Under this assumption, equation~\ref{eq:argmin_full} simplifies to: 
\begin{equation}
    \label{eq:argmin_simple}
    \defMisreg^{\,*}= \argmin_{g, \defMisreg}^{\,} \left\| \hat{\mathbf{D_\alpha}} -  \begin{bmatrix}\mathbf{D_{\alpha_0}}, & \mathbf{\Lambda_{\alpha_0}}\end{bmatrix} \cdot \begin{bmatrix} g \\ g\defMisreg\end{bmatrix} \right\| 
\end{equation}
which is of the form $\left\| \mathbf{Y} -  \mathbf{X} \mathbf{\beta} \right\|$ and can therefore be solved quickly using a standard least squares solver.
Since the linear assumption used to produce the meta sensitivity matrix only holds for a local working point around the registration $\alpha_0$, we will need to perform several cycles of updating the derived misregistration, $\alpha_i$, and then recomputing a new meta sensitivity matrix about this new point before computing $\alpha_{i+1}$. 
This iterative solving for the misregistion vector is analogous to gradient descent, and as such, we introduce a learning rate vector, $\vec{\gamma}$, and update our observed misregistrations as follows:
\begin{equation}
    \defMisreg_{i+1} = \defMisreg_i + \vec{\gamma}\cdot\defMisreg^*_i
\end{equation}
Once the change in the derived optimal misregistration approaches zero, the solution is said to have converged and a best-fit pseudo-synthetic IM is computed.
Furthermore, the learning rate can be tuned to the system to increase the speed of convergence by examining the shape of the convergence graph (Figure~\ref{fig:convergence}).

While Equation~\ref{eq:argmin_simple} is defined for the entire IM, we note that the derivation made no assumption regarding the number of modes in the matrix.
In practice, we are able to drastically speed up our solver by running the algorithm in batches of only a few modes (we have found that $\sim 5$ modes works well).
This approach saves time by not simulating the entire IM at each step in the convergence.
Furthermore, as discussed in the original SPRINT paper, it is possible to only use a set of specific modes with are highly sensitive to misregistrations, although this was not attempted in this work.

\section{LAB RESULTS}
\label{sec:lab} 

In this section, we will show validation of our methodology conducted on the Renovated Adaptive Z-band Optic Relay system (RAZOR) AO testbed located at the University of Toronto's Spectroscopic High Angular Resolution/Photonics (SHARP) imaging lab.
Using RAZOR, we are able to directly compare the performance of the ESCAPE calibration method to the standard IM calibration for both Shack-Hartmann and Pyramid wavefront sensors. 

\begin{figure} [ht]
\begin{center}
\begin{tabular}{c}
\includegraphics[width=0.9\textwidth]{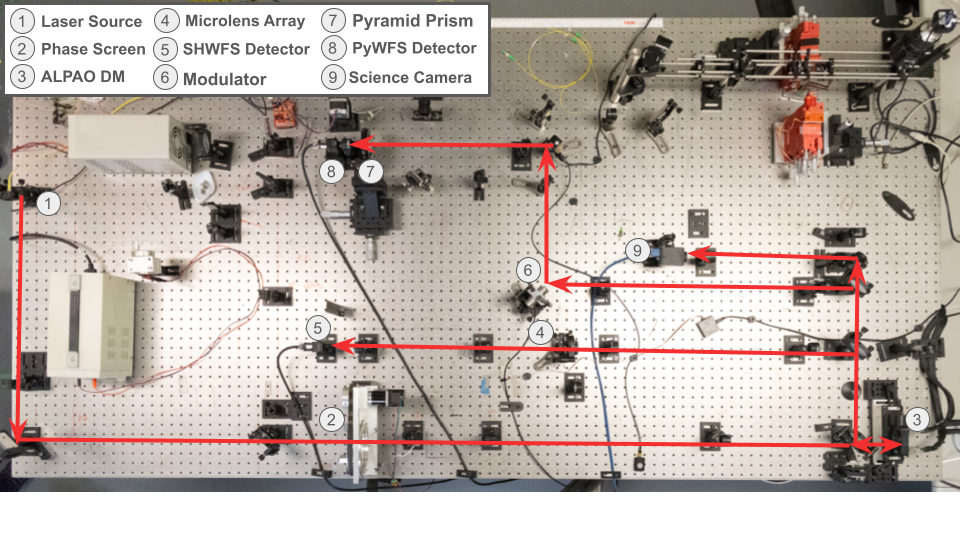}
\end{tabular}
\end{center}
\caption[example] 
{ \label{fig:razor} A photo of the RAZOR AO testbed from above with the relevant light path highlighted in red. The system is illuminated by a 635nm red laser source (1). Next, the light passes through one of four possible turbulent phase screens  placed directly before the aperture (2). The turbulent wavefront then reflects off of an ALPAO DM-97 (3), before being split and sensed by the SHWFS arm (4-5) or the Pyramid arm (6-8). The focal plane PSF is imaged by a FLIR Grasshopper3 camera (9).}
\end{figure} 

\subsection{Experimental Set-Up}

In its current configuration, RAZOR (see Figure ~\ref{fig:razor}) supports two co-aligned sources; one is a standard red laser (635nm), and the other is an infrared LED which illuminates an extended object.
All experiments reported in this work were illuminated with the 635nm laser source.
All optics used on the bench are achromatic from visible to NIR to facilitate simultaneous optical and NIR operation.
Both sources pass through at $6.75$mm circular aperture and a simulated atmosphere --- one of four possible rotating phase screens with $r_0$ between $0.2$ and $1.5 mm$ --- before illuminating the surface of a 97 actuator ALPAO deformable mirror (DM97-15).
After reflecting off of the DM97, the light is separated into two independent wavefront sensing arms: a SHWFS arm, and a PyWFS arm.
The SHWFS arm uses an off-the-shelf microlens array (MLA) and a XIMEA CMOS camera (MQ013MG-ON).
The PyWFS arm is composed of a PBL25Y/S-BAM12 glass Pyramid, a tip-tilt modulator (PI S-330 Tip Tilt and E-727 Controller), and a second XIMEA CMOS camera (MQ013MG-ON). 
The light not picked off by the sensing arms passes through to RAZOR's `science' camera (Teledyne Flir Grasshopper 3). 

RAZOR is controlled using the open-source AO real-time control (RTC) software \href{https://github.com/jacotay7/pyRTC}{pyRTC}\footnote{For source code, see: https://github.com/jacotay7/pyRTC}, a package originally developed to control RAZOR.
pyRTC is an open-source Python package for real-time control of adaptive optics (AO) systems, aimed at making high-performance AO control more accessible to the AO community at large.
It provides an efficient RTC pipeline, supporting operation both as a single process with multithreaded control tasks, or as network of coordinated Python processes which communicate via shared memory and TCP sockets.
Key features include community-driven development, cross-platform compatibility, and a growing library of examples interfacing with common AO hardware.
We encourage all interested members of the community to try out pyRTC and contribute to the growing codebase.

\subsubsection{IM Performance Validation Procedure}
\label{sec:validation}
Given a candidate IM for an AO system, there are several hyper-parameters which can drastically impact the performance of the control loop.
For instance, a given IM will have an optimal control gain under a given atmospheric condition, and the performance will vary sharply as this gain is modified.
Other parameters which can affect loop performance include: control speed, wind speed, $r0$, number of controlled modes used, leaky gain, integrator specific parameters, etc...
As the complexity of the control system increases, so does the number of parameters to optimize.
When comparing calibrations taken with differing methodologies, the specific set of hyper-parameters which result in the optimal performance for one method might not be optimal for the other. 
In particular, when trying to prove that performance is equivalent or only marginally improved, the hyper-parameters chosen can often have a larger impact that the calibration used.
This is particularly relevant for synthetic matrices, which we have empirically observed can have significantly different optimal control variables.
This is a subtle, but extremely important constraint on our testing; if we were to fix the hyper-parameters between matrices, our results would be extremely biased by those choices.
To combat this, we have implemented automated hyper-parameter sweeps using the \href{https://wandb.ai/site}{Weights \& Biases} (WandB) platform.

WandB sweeps are a powerful tool for optimizing hyper-parameters, usually employed in the context of training machine learning models. 
In the context of comparing the closed-loop performance between two calibration matrices, we define an experiment such that the sweep is able to adjust all of the loop hyper-parameters, including which IM is used.
Then, the sweep will parse the hyper-parameter space and evaluate the \textit{importance} of each parameter relative to a given performance metric, in our case the Strehl Ratio (SR).
If the choice of calibration matrix has little or no effect on the performance metric (i.e. low importance), then we can conclude the control matrices are effectively equivalent. 
Alternatively, if one matrix is statistically out performing the others across a wide set of hyper parameters we can conclude it is superior.
This procedure removes the need for manual hyper-parameters tuning, and provides a definitive procedure for comparing matrices which are optimal under differing conditions.

\subsection{Lab Performance}

Since the primary objective of the ESCAPE method is to calibrate an AO system which does not possess an internal calibration source, the method can be said to have been successful if the resulting calibration matrix performs at least as well as the baseline calibration.
As such, we do not aim to provide a method which is in some way an improvement over traditional methods, but rather provide a methodology which is both simple and effective for systems without access to a calibration source.
In this work, we will examine the performance of the ESCAPE method using both a Shack-Hartmann WFS and a Pyramid WFS. 
In each case, we will compare the baseline calibration method to a DO-CRIME calibration, as well as a pseudo-synthetic matrix derived from a  DO-CRIME matrix (an ESCAPE matrix).
For the PyWFS, in addition to the above procedure, we will also use a DO-CRIME matrix taken in closed-loop to provide an optical gain vector.
We ran the SHWFS experiments at a loop speed of 500Hz, while the PyWFS tests were run at 300Hz; the phase screen speed was adjusted to mostly compensate for the differing loop speeds between the WFSs.
Both loops were closed using a standard leaky integrator and a Lexitech phase screen ($D/r_0 \sim 22)$ rotating at constant speed.
All PSFs presented were taken in the visible, illuminated by a 635nm laser source.

\subsubsection{Shack-Hartmann Wavefront Sensor}
\label{sec:lab_shwfs}

\begin{figure} [ht]
\begin{center}
\begin{tabular}{c}
\includegraphics[trim={0cm 3cm 0cm 3cm},clip, width=0.95\textwidth]{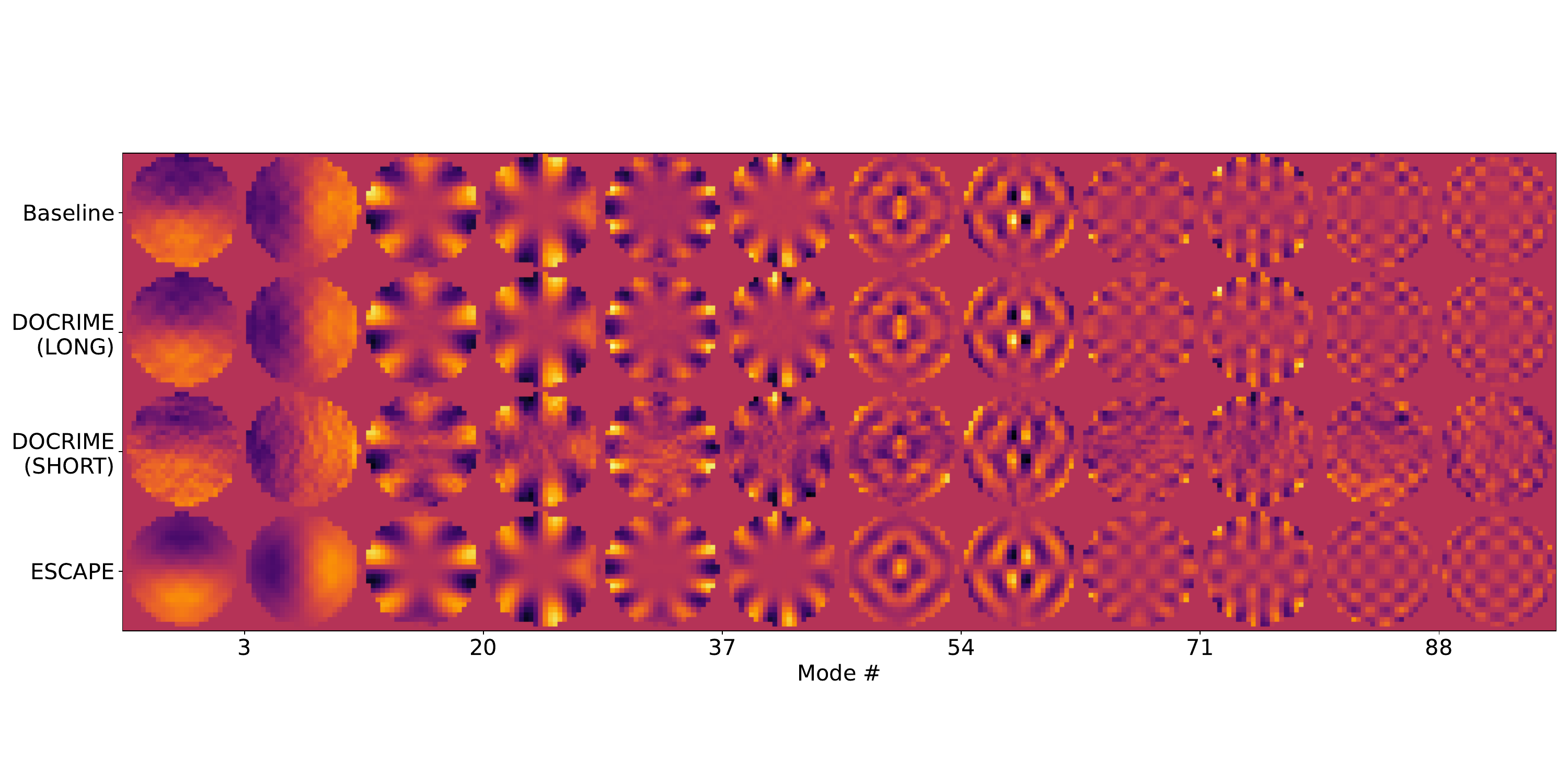}
\end{tabular}
\end{center}
\caption[example] 
{A visual comparison of the IMs examined in the SHWFS study presented in this work. For compactness, an evenly spaced set of representative modes have been selected from the matrices for examination. As our IM was made from slope maps, each mode contains the X and Y slopes displayed as a pair. All of the modes were normalized to be more easily visualized on the same colorbar. The \textit{baseline} matrix was created using the standard push-pull method on a calibration source. The DO-CRIME matrices were created in open-loop using the methodology described in Sec~\ref{sec:DO-CRIME}. We present the matrices from both the SHORT and LONG experiments described in Sec~\ref{sec:lab}. Lastly, the ESCAPE matrix is a pseudo-synthetic matrix which was created using the DO-CRIME matrices and our OOPAO simulation.}
\label{fig:SHWFS_IM_COMP}
\end{figure}

In order to demonstrate the effectiveness of ESCAPE on our SHWFS, we present two experiments which differ only in the number of sky measurements used to compute the open-loop DO-CRIME matrix.
For our first experiment (hereafter, the LONG experiment), we ran enough loop cycles for the DO-CRIME matrix to show high signal-to-noise in all modes.
An arbitrary cutoff of 100,000 cycles was used ($\approx 3$ minutes at 500Hz).
For our second experiment (hereafter, the SHORT experiment), we repeated the same with only 10,000 cycles ($\approx 20$ seconds at 500Hz).
Next, both of the DO-CRIME matrices were fed into a pseudo-synthetic IM solver based on the SPRINT method (see Sec \ref{sec:SPRINT}).
For our implementation of SPRINT, we simulated the RAZOR bench using the OOPAO\cite{OOPAO} AO simulation package and  wrote a custom implementation of the SPRINT procedure following the methodology shown in Sec~\ref{sec:SPRINT}.
Since a pseudo-synthetic matrix can be fully described by the misregistration vector, $\defMisreg$, it is sufficient to show that if we can recover the same misregistration vector for both experiments, then the ESCAPE calibration will be identical.
This experiment highlights one of the most prominent strengths of using a pseudo-synthetic technique for systems lacking a calibration source; by reducing the dimensionality of the problem, we are able to more effectively utilise limited and noisy calibration measurements.
Table~\ref{tab:misreg} presents the inferred misregistration parameters from both experiments, showing that we are able to consistently recover an equivalent misregistration vector, and therefore produce equivalent pseudo-synthetic matrices using either the SHORT or LONG DO-CRIME inputs.
Figure~\ref{fig:SHWFS_IM_COMP} shows a visual comparison of all of the IMs used for this analysis. 

\begin{table}[ht]
    \centering
    \begin{tabular}{ccc}
        \toprule
        Component & \(\alpha_{\text{SHORT}}\) & \(\alpha_{\text{LONG}}\)\\
        \midrule
        Shift X [\% of Sub Aperture] & $-15.3$ & $-16.6$\\
        Shift Y [\% of Sub Aperture] & $-71.3$ & $-71.7$ \\
        Rotation [deg]&$-0.75$ & $-0.77$\\
        Radial Scaling [\% of Diameter] & $2.9$ & $2.9$ \\
        Tangential Scaling [\% of Diameter] & $0.17$ & $-0.02$\\
        \bottomrule
    \end{tabular}
    \\
    \caption{A comparison between misregistration vectors determined from running our SPRINT pseudo-synthetic matrix routine on two input IMs. The misregistration vector, $\alpha_{\text{SHORT}}$, corresponds to the output from the routine when it was fed a DO-CRIME matrix corresponding to $\approx 20$ seconds of data. The misregistration vector, $\alpha_{\text{LONG}}$, corresponds to the output from the routine when it was fed a DO-CRIME matrix corresponding to $\approx 3$ minutes of data. Ignoring tangential scaling (which was consistent with 0), all misregistration parameters were consistent to within a maximum error of 8\% for this experiment. These results show that we are able to measure consistent misregistrations (and therefore produce effectively equivalent pseudo synthetic solutions) using relatively noisy matrices which would otherwise be unable to provide optimal performance themselves.}
    \label{tab:misreg}
\end{table}

\begin{figure} [ht]
\begin{center}
\begin{tabular}{c}
\includegraphics[width=\textwidth]{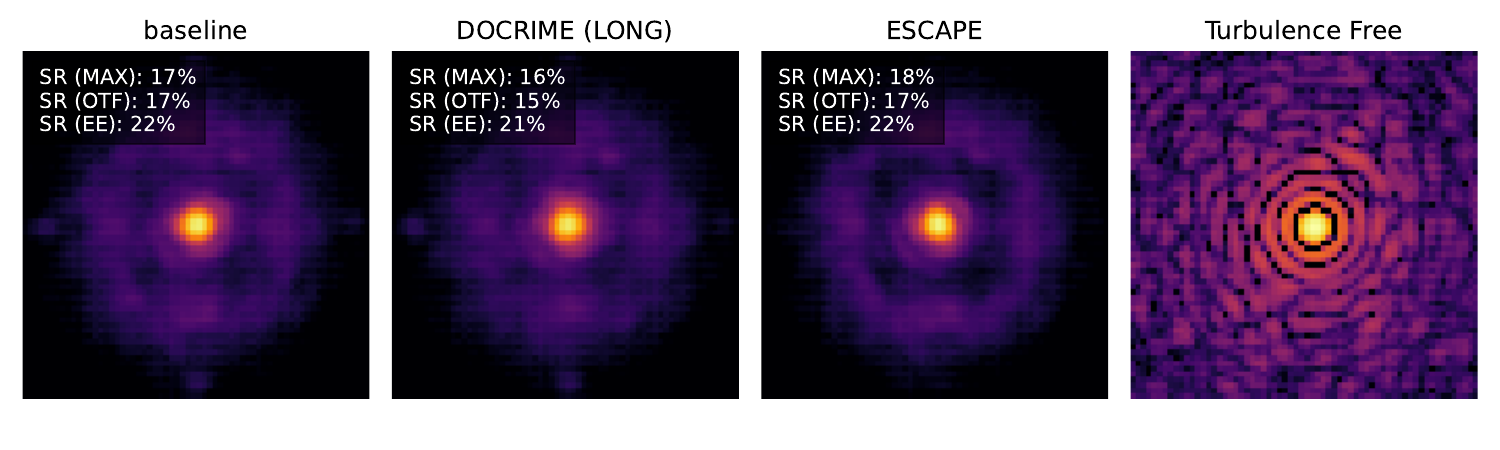}
\end{tabular}
\end{center}
\caption[example] 
{Stacked closed-loop PSFs from each of the three calibration matrices examines in our experiment and the corresponding empirical model taken without turbulence. For completeness, we present three method for determining the Strehl ratio from the images. In order from top to bottom, the methods are comparing the max value of the image to the model, comparing the optical transfer functions of the PSFs to the model, and lastly, comparing the enclosed energy in a 3 pixel radius of the maximum. A radius of 3 pixels was chosen to match the approximate distance to the first minimum. }
\label{fig:SHWFS results}
\end{figure}

To test the performance of our calibrations, we followed the validation procedure outlined in Sec~\ref{sec:validation} to control for the differing preferred hyper-parameters of each matrix.
In total, we ran 1,000 randomized trials varying the calibration matrix used as well as the number of dropped modes, loop gain, and leaky gain. 
We note that each matrix did have slight differences in optimal hyper-parameters which could easily have led us to dismiss one or the other as performing worse if not all parameters were controlled.
Overall, the matrices performed equivalently to within the sensitivity of our analysis.
We conclude that, at least for our tests, the ESCAPE method is a viable method for calibration without a calibration source.
Figure~\ref{fig:SHWFS results} presents the stacked point-spread functions (PSFs) for each of the three matrices investigated in the sweep with their optimized hyper-parameters, the Strehl ratios are consistent between the baseline and ESCAPE matrices, with the DO-CRIME performing very slightly worse.
We note that a major source of error on our bench are faint vibrations which cause erroneous tip-tilt correction in the AO loop; because of this, changes in Strehl on the order of few percent are easily produced by vibrations.

\subsubsection{Pyramid Wavefront Sensor}
\label{sec:lab_pywfs}

PyWFS are known to have a significantly reduced linear regime compared to their SHWFS counterparts\cite{Pyramid_Linearity}.
Commonly, this effect is offset through the use of modulation, which  effectively trades sensitivity for a larger linear range; however, in open-loop, aberrations are still large enough to be comfortably outside the linear regime of the pyramid.
Therefore, the on-sky calibration of PyWFSs is distinctly different than SHWFSs.
On-sky calibrations are, by definition, taken on top of uncorrected atmospheric turbulence; therefore, the magnitude of the observed WFS response will be distinctly different than when measured around a turbulence-free calibration source.
This effect is equivalent to the optical gain (OG) problem which has been extensively investigated in the literature\cite{optical_gain_comp}.
In this case, instead of compensating for the (relatively) small difference between the pyramid sensitivity in closed-loop versus around a calibration source, we instead need to compensate for the much larger difference between open-loop and closed-loop conditions.
Overall, this effect results in on-sky IM's  having optical gain coefficients greater that 1.

To solve this problem, we propose utilizing closed-loop DO-CRIME calibrations with successively lower amplitude pokes in order to bootstrap towards the optimal optical gain vector.
The fundamental idea is that an on-sky IM taken in closed-loop will directly measure the local linear approximation of the sensor around the desired operating turbulence, and therefore, be used to extract the optical gain vector.
Specifically, for a given mode, we scale the standard deviation of the mode to match the closed-loop IM. Or equivalently:
\begin{equation}
    \vec{OG}_i = \frac{stdev(\defIM_i)}{stdev(\widehat{\defIM_i})}
\end{equation}
where $i$ denotes the $i^{th}$ mode, $\defIM$ is the true IM around the current turbulence, and $\widehat{\defIM_i}$ is our IM taken around a different turbulence condition.
The re-scaled IM is then found simply with $\vec{OG}\cdot\widehat{\defIM_i}$.
Our full procedure proceeds as follows.
Initially, a high signal-to-noise ratio (SNR)  DO-CRIME calibration is taken and is used to produce a pseudo-synthetic calibration matrix.
Next, the loop is closed on either the pseudo-synthetic calibration or the open-loop DO-CRIME matrix; this step can be particularly challenging since the initial calibration scaling may be sufficiently incorrect such that the loop does not close stably.
Once the loop is closed, a closed-loop DO-CRIME matrix is taken with minimal amplitude randomized pokes ($\sim 0.1-0.2\%$ of the DM stroke).
The produced matrix is used to extract a direct measurement of the OGs since it was taken around the closed-loop conditions.
The high SNR matrix is then updated with the measured OGs and the process is repeated several times, reducing the amplitude of the DO-CRIME pokes at each step until convergence.

\begin{figure} [ht]
\begin{center}
\begin{tabular}{c}
\includegraphics[trim={5.5cm 0cm 8cm 0cm},clip, width=0.95\textwidth]{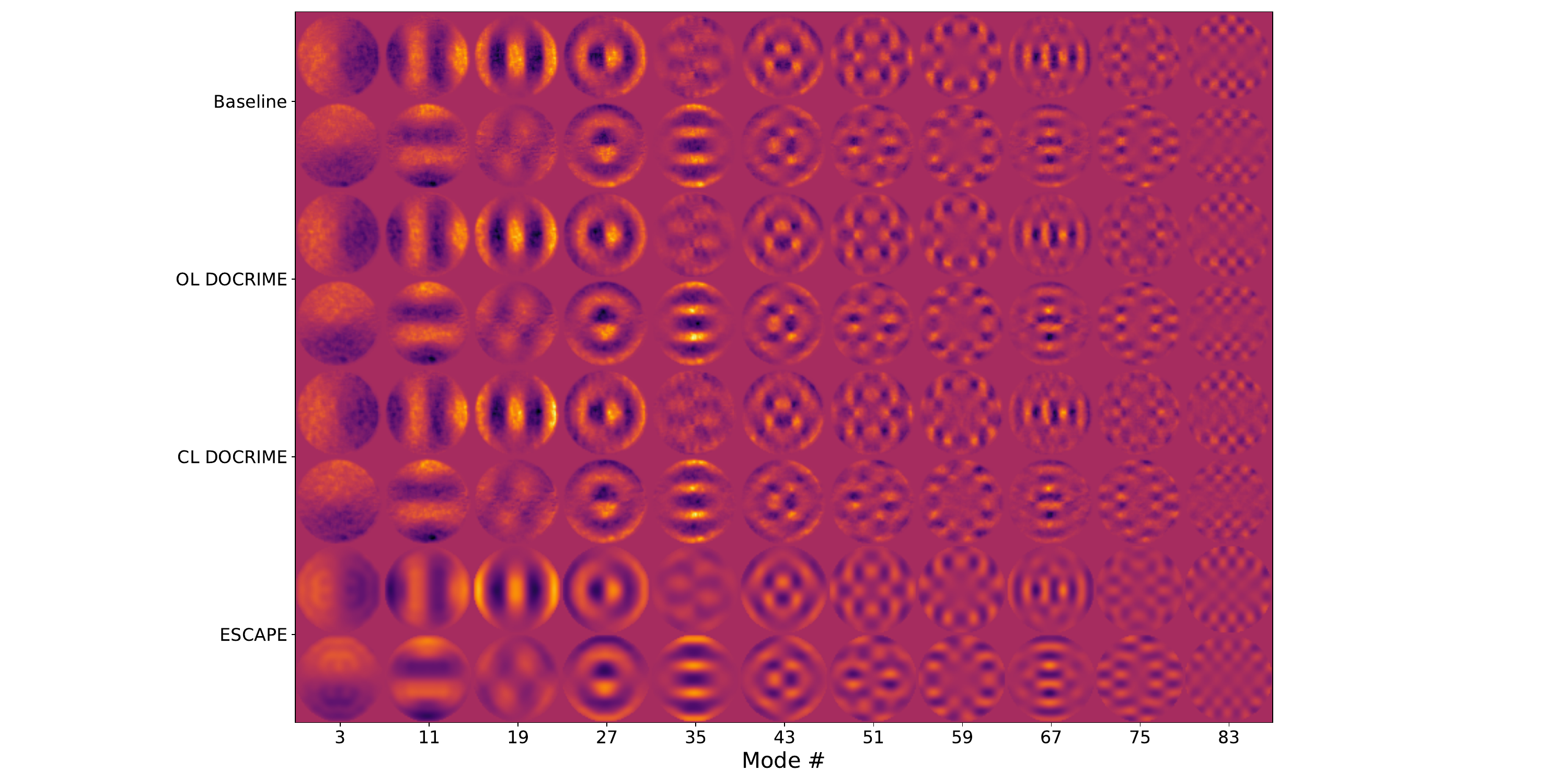}
\end{tabular}
\end{center}
\caption[example] 
{A visual comparison of the IMs examined in the PYWFS lab study presented in this work. For compactness, an evenly spaced set of representative modes have been selected from the matrices for examination. As our IM was made from slope maps, each mode contains the X and Y slopes displayed as a pair. All of the modes were normalized to be more easily visualized on the same colorbar. The \textit{baseline} matrix was created using the standard push-pull method on a calibration source. The DO-CRIME matrices were created using the methodology described in Sec~\ref{sec:DO-CRIME}. Lastly, the ESCAPE matrix is a pseudo-synthetic matrix which was created using the open-loop DO-CRIME matrix and our OOPAO simulation.}
\label{fig:PYWFS_IM_COMP}
\end{figure}

\begin{figure} [ht]
\begin{center}
\begin{tabular}{c}
\includegraphics[trim={0cm 0cm 0cm 0cm},clip, width=0.95\textwidth]{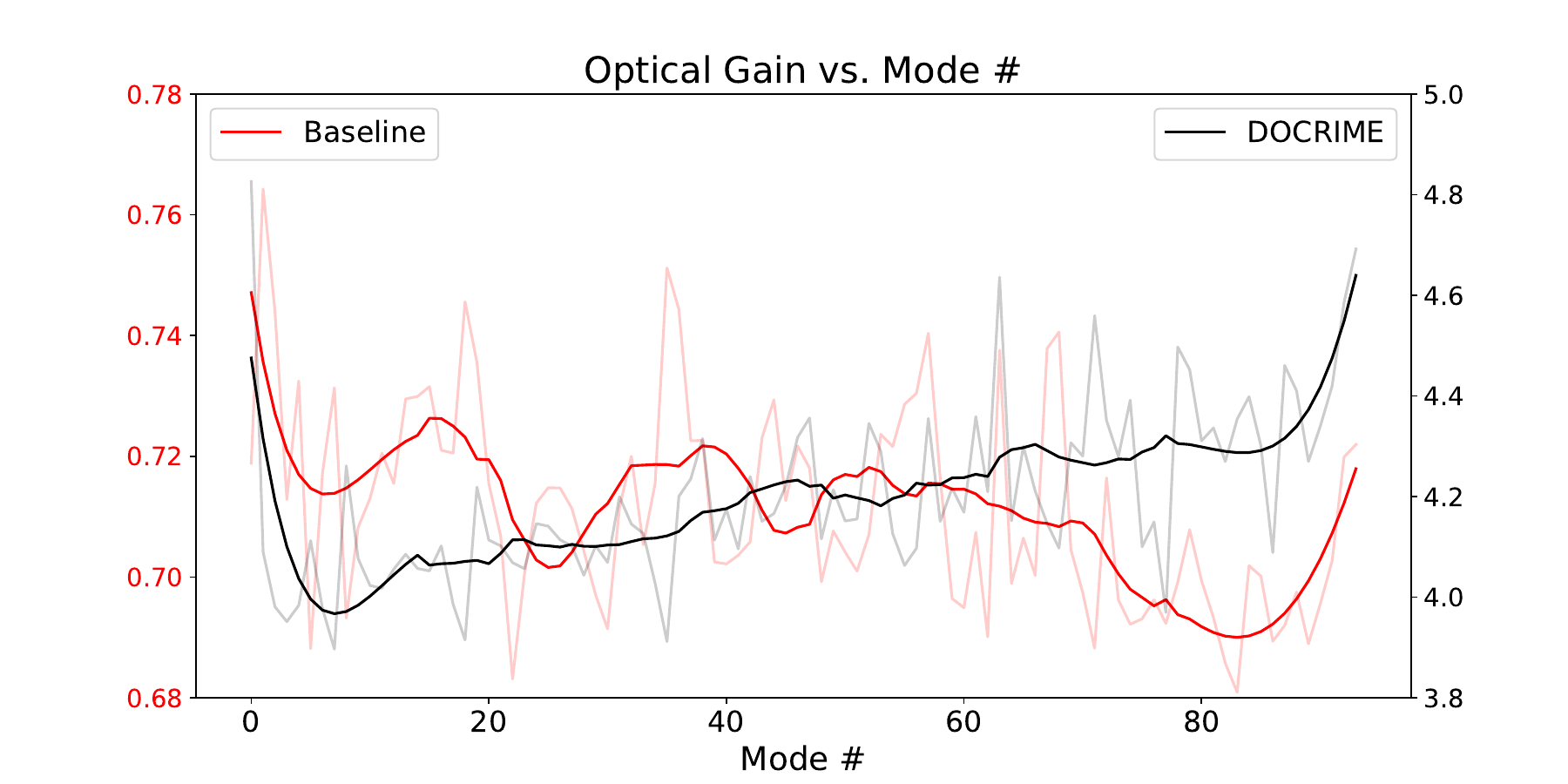}
\end{tabular}
\end{center}
\caption[example] 
{The optical gain vectors found for both the baseline IM (red) and the DO-CRIME IM (black). The vector values are plotted in addition to a smoothed overlay to show the general trends. Note that there are two y-axes, the left axis corresponds to the values for the baseline IM, while the right axis corresponds to the DO-CRIME IM. Because the baseline IM was taken around a calibration source, we expect the optical gain vector to be less than 1, while we expect the optical gain vector for the DO-CRIME IM (taken in open-loop) to be greater than 1.}
\label{fig:optical_gains}
\end{figure}

\begin{figure} [ht]
\begin{center}
\begin{tabular}{c}
\includegraphics[trim={0cm 0cm 0cm 0cm},clip, width=0.95\textwidth]{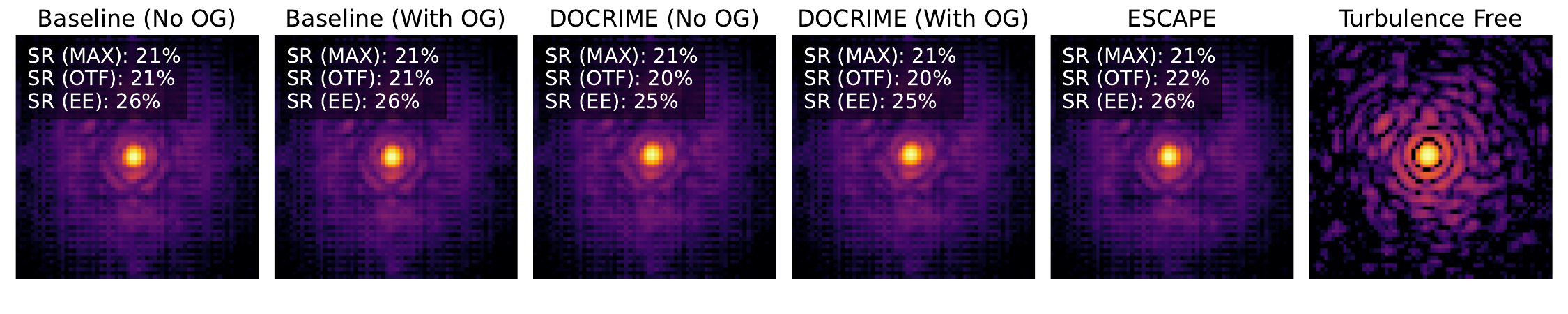}
\end{tabular}
\end{center}
\caption[example] 
{Stacked closed-loop PSFs from each of the three calibration matrices examined in our experiment, both with and without OG scaling, as well as the corresponding empirical model taken without turbulence. For completeness, we present three method for determining the Strehl ratio from the images. In order from top to bottom, the methods are comparing the max value of the image to the model, comparing the optical transfer functions of the PSFs to the model, and lastly, comparing the enclosed energy in a 3 pixel radius of the maximum. A radius of 3 pixels was chosen to match the approximate distance to the first minimum. We note that across several experiments, we see very comparable performance, with a marginal performance preference towards the pseudo-synthetic calibration.}
\label{fig:PYWFS_ESCPAE_PSFs}
\end{figure}

In order to validate this procedure, we performed a full test on the RAZOR test bench.
In total, our procedure produced four interaction matrices (Figure~\ref{fig:PYWFS_IM_COMP}): the baseline matrix taken using the standard calibration method, an open-loop DO-CRIME matrix (100,000 iterations), a closed-loop DO-CRIME matrix taken while the loop was closed with the open-loop DO-CRIME matrix, and an ESCAPE matrix made by feeding SPRINT with the open-loop DO-CRIME matrix.
The empirical matrices (both the baseline and DO-CRIME) were scaled to match the closed-loop DO-CRIME matrix, effectively applying our determined optical gain vector.
We tested the matrices with and without the optical gain scaling.
We note that, for our testing conditions, the optical gains we found were roughly constant across mode (Figure~\ref{fig:optical_gains}).
This meant that the optical gain scaling could be accomplished by simply adjusting the loop gain, resulting in very similar performance across the matrices, but with drastically different loop hyper-parameters.
Following our IM validation procedure (Sec~\ref{sec:validation}), we ran a WandB hyper-parameter sweep for a case of 1000 different loop parameter configurations, both with and without applying the OG vector.
Figure~\ref{fig:PYWFS_ESCPAE_PSFs} presents the stacked PSF images and observed Strehl ratios for all of the tested calibration matrices under their optimized loop parameters.
We observe an identical results to the case of the SHWFS, where the baseline and ESCAPE method perform similarly and the DO-CRIME matrix alone is only very marginally worse.
We note that in both cases, and throughout all of our experimentation, the pseudo-synthetic matrix does produce a slightly higher contrast control radius.

\section{ON-SKY PROGRESS}

\label{sec:sky}  

\subsection{MAPS}

The MMT Adaptive optics exoPlanet characterization System (MAPS) is an exoplanet characterization mission which consists of upgrades to both the MMT's AO system and AO-fed science instruments. 
The MAPS AO system upgrade includes a refurbished 336-actuator adaptive secondary mirror (ASM), two pyramid wavefront sensors (PyWFS's), and updated control software. 
The MAPS program also includes substantial upgrades to both Arizona InfraRed Imager and Echelle Spectrograph (ARIES) and the MMT-sensitive polarimeter (MMT-Pol) science instruments.
These instruments will support a 60-night observational campaign to study the atmospheric composition and dynamics of 50-100 exoplanets. 
Currently, MAPS has been undergoing on-sky commissioning since Fall 2022, with most efforts relating to the commissioning of the AO system.
A thorough update on the current commissioning progress, setbacks, and status can be found in Morzinski et al. (2024).

As mentioned repeatedly in this work, one of the unique difficulties when commissioning AO systems with pre-focal, convex ASMs is performing system calibration tasks without access to a calibration source.
For this reason, calibration was identified as one of the key areas of concern for the MAPS project, sparking the investigation and development of ESCAPE.
In parallel to the MAPS commissioning efforts, we have been developing simulations and control code in order to reproduce our lab results and methodologies (Sec~\ref{sec:lab_pywfs}) on a facility-class AO system.
In this section, we will provide a comprehensive update on our best efforts to date, with some descriptions of the current hurdles we are working to overcome.

\begin{figure} [ht]
\begin{center}
\begin{tabular}{c}
\includegraphics[trim={8.5cm 0cm 11.5cm 0cm},clip, width=0.95\textwidth]{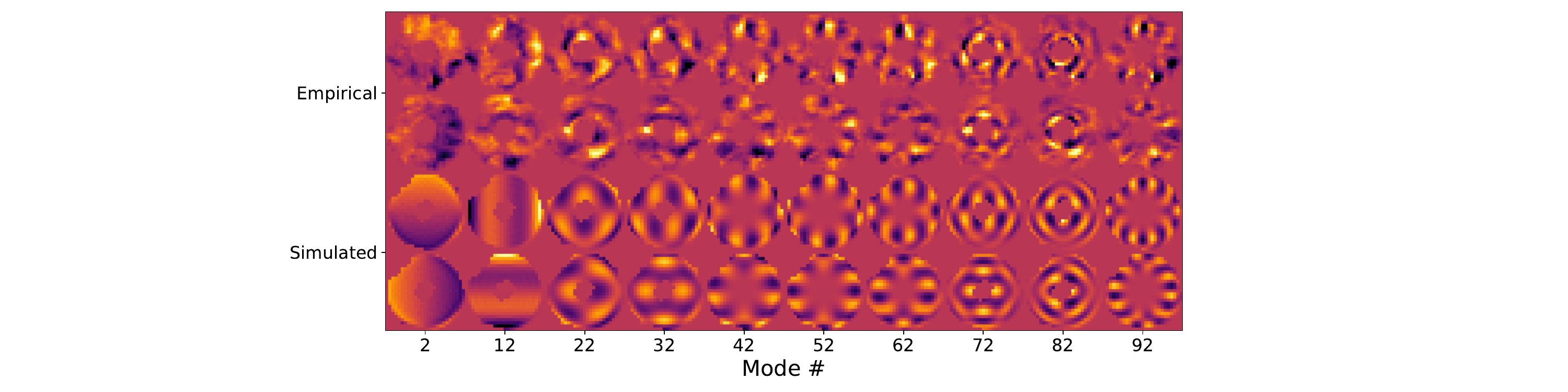}
\end{tabular}
\end{center}
\caption[example] 
{A 100 mode DO-CRIME IM taken on MAPS using the near-infrared pyramid WFS compared to a simulated IM from our OOPAO simulation of MAPS. The are several areas missing from the side of the pupil due to both contamination and poor illumination resulting from incorrectly positioned edge actuators. We observe non-trivial signal in all of the modes, however there are significant asymmetries and missing signal in the modes. This points to either an issue the ASM's responsiveness to a given mode, or issues with illumination of the subapertures under high turbulence. We note that all of these challenges would be encountered by any on-sky calibration methodology and the DO-CRIME method remains fast and robust. }
\label{fig:MAPS_IM}
\end{figure}

During our May 2024 run, we dedicated 2 nights to the testing of ESCAPE on-sky.
In order to fully utilise the code that has been previously developed in the lab, and to provide a more apt comparison of performance, we interfaced the nominal MAPS control code directly into pyRTC for this testing.
Since MAPS uses two PyWFS (one infrared, one visible), we attempted to follow an analogous procedure to the one taken in Section~\ref{sec:lab_pywfs}.
The first issue encountered was contamination in the mirror; since the ASM magnetically floats slightly above the surface of its reference body ($\approx 50$um), it is prone to contamination at the edge of the mirror.
Contamination is an ongoing issue with the MAPS ASM, and often, can be resolved during a commissioning run.
During this particular run, the contamination could not be removed; because of this, several of the actuators around the contamination were turned off, and therefore, a section of the ASM will appear to be missing in the IM.
A second issue is general illumination of the edge of the ASM.
During this run, several edge sections of the mirror were not flat enough to properly fully illuminate the pupil in all sub apertures.
This effect, combined with the contamination, cause the pupil to be non-circular around its edge, which can be seen in the IM.
Because of issues in the pupil quality, we did not attempt to produce a pseudo-synthetic matrix of the system, and instead focused solely on producing an open-loop DO-CRIME matrix.
Figure~\ref{fig:MAPS_IM} presents a 100 mode open-loop DO-CRIME IM taken on MAPS.

Through these tests, we have noted that the DO-CRIME approach is able to measure on-sky IMs quickly, taking only a few minutes of sky-time to build up significant signal in all modes.
Once the system has developed enough (or our simulation has developed enough) to meet the criteria of SPRINT, we will be able to rapidly produce ESCAPE matrices. 
One consistent issue we have run into is the limited dynamic range of the PyWFS.
This is particularly relevant when taking on-sky IMs which require introducing signals on top of an existing atmospheric signal.
Our initial plan was to first close the loop on an open-loop DO-CRIME matrix before taking a subsequent close-loop DO-CRIME matrix, but we were not able to accomplish that during this run.
For future tests, we are planning on further increasing the modulation radius to try and obtain cleaner open-loop matrices prior to producing a closed-loop matrix.

\section{CONCLUSION}
\label{sec:conclusion} 

In this work, we have detailed a calibration procedure for AO systems lacking a calibration source which combines both existing empirical and pseudo-synthetic methodologies.
Namely, our ESCAPE method is a direct combination of the DO-CRIME \& SPRINT techniques, with some additional minor customization.
We have presented laboratory results from the RAZOR test bench, located in the SHARP lab at the University of Toronto, which have shown that ESCAPE can perform equivalently to the baseline calibration method for both SHWFSs and PyWFSs.
Moreover, we have presented a procedure for using closed-loop DO-CRIME calibrations as a method of OG measurement and compensation, which has the potential to be applied to any PyWFS system. 
Lastly, we have detailed our current best efforts to apply this technique on-sky using the MAPS AO system on the MMT.

In future works, we aim to fully test this methodology using an on-sky system; namely, our plan is to use the MAPS AO system as a full scale test of this methodology in a facility-class AO system.
Also, we have already begun work to retrofit the REVOLT AO system\cite{REVOLT}, located at the Dominion Astrophysical Observatory, with a pyRTC-based control system.
Running on REVOLT will allow us to fully test the ESCAPE methodology on-sky for both a SHWFS and PyWFS with direct comparison to a standard calibration baseline.

\acknowledgments       

We would like to thank the entire MAPS team and MMT staff for their continued and invaluable work on the on-going commissioning efforts.

The MAPS project is primarily funded through the NSF Mid-Scale Innovations Program, programs AST-1636647 and AST-1836008. 

The near infrared WFS contributions to MAPS have been funded by the Canada Foundation of Innovation and the Ontario Research Funds.

The work of authour J.T. is supported by the Ontario Graduate Scholarship (OGS) program and the FAST doctoral fellowship program at the University of Toronto.

\bibliography{main}
\bibliographystyle{spiebib} 

\end{document}